\documentclass{svjour2}
\usepackage{graphicx}
\usepackage{latexsym}
\usepackage{amssymb}

\def\beq{\begin{equation}}
\def\eeq{\end{equation}}
\def\eqref#1{(\ref{#1})}

\journalname{General Relativity and Gravitation}

\begin{document}

\title{New solutions of the Ermakov-Pinney equation in curved space-time}

\author{Donato  Bini \and  Giampiero Esposito}

\institute{
Donato Bini
\at
Istituto per le Applicazioni del Calcolo ``M. Picone,'' CNR, I--00161 Rome, Italy\\
Istituto Nazionale di Fisica Nucleare, Sezione di Napoli, \\ 
Complesso Universitario di Monte S. Angelo,
Via Cintia Edificio 6, 80126 Naples, Italy\\ 
ORCID: 0000-0002-5237-769X\\
\email{donato.bini@gmail.com}
\and
Giampiero Esposito 
\at
Istituto Nazionale di Fisica Nucleare, Sezione di Napoli, \\
Complesso Universitario di Monte S. Angelo, Via Cintia Edificio 6, 
80126 Napoli, Italy\\
ORCID: 0000-0001-5930-8366\\
\email{gesposit@na.infn.it}
}

\date{Received: date / Accepted: date / Version: \today}

\maketitle

\begin{abstract}
An Ermakov-Pinney-like equation associated with the scalar wave equation in curved 
space-time is he\-re stu\-di\-ed. The ex\-am\-ple of 
Schwarzschild spa\-ce-ti\-me con\-si\-de\-red in 
the pre\-sent wo\-rk shows that this equation can be viewed more as a \lq\lq model 
equation," with interesting applications in black 
hole physics. Other applications studied involve cosmological spacetimes (de Sitter) 
and pulse of plane gravitational waves: in all these cases the evolution of 
the Ermakov-Pinney field seems to be consistent with a rapid 
blow-up, unlike the Schwarzschild case where 
spatially damped oscillations are allowed.
Eventually, the phase function is also evaluated in many of the above spacetime models.
\end{abstract}
 

\maketitle

\section{Introduction}

Let $({\mathcal M},g)$ denote a four-dimensional space-time with metric $g$ 
of Lorentzian signature $+2$, and 
with associated scalar product defined as 
$$
\langle A,B \rangle \equiv g_{\mu\nu}A^\mu B^\nu=g(A,B).
$$
The scalar wave equation describing a massless scalar field 
propagating on  $({\mathcal M}, g)$ reads
\begin{equation}
\label{eq:1}
\Box \chi=0 ,
\end{equation}
where the \lq\lq box operator" is the familiar wave operator in curved space-time
\begin{equation}
\Box= g^{\mu\nu}\nabla_\nu \nabla_\mu . 
\end{equation}
Equation \eqref{eq:1} has been largely studied in the literature, in various contexts including black 
holes, cosmological spacetimes, gravitational wave spacetimes, etc.
In particular, as shown in Ref. \cite{CK}, the technical difficulty of the coupled 
nature of Maxwell equations in curved space-time can be overcome by mapping them into a wave
equation for a complex scalar field. The real and imaginary part of such a field are 
therefore ruled by a scalar wave equation as Eq. \eqref{eq:1}. This property is
eventually applied to the investigation of binary systems in relativistic 
astrophysics \cite{CK}. 

The standard approach to Eq. \eqref{eq:1} is the separation of variables when the symmetries of 
space-time allow for it, or mode-sum decomposition when a Fourier analysis can be performed.
In general, the theory developed in Ref. \cite{Treves} suggests looking for a solution $\phi$ which,
up to a remainder term, consists of a function $\Phi$ obtained by integrating products
of functions depending on all cotangent bundle local coordinates. This occurs because Fourier
transforms must be replaced by Fourier-Maslov integral operators on passing from 
Minkowski space-time to curved pseudo-Riemannian manifolds \cite{Treves}. 
However, we here point out that, since the scalar wave equation is linear, 
one can always consider 
\begin{equation}
\chi=\alpha e^{i \varphi},
\end{equation}
where $\alpha$ and $\varphi$ are real-valued, with the understanding
that ${\rm Re}(\chi)=\alpha \cos \varphi$ is the desired solution 
of the scalar wave equation with variable real-valued coefficients.
One therefore finds the coupled set \cite{EBD,EM} of equations (here written first in
dimensionless units for simplicity)
\begin{equation}
{\rm div}(\alpha^{2}{\rm grad}\varphi)\equiv \nabla^\mu (\alpha^2 \nabla_\mu \varphi)=0,
\end{equation}
\begin{equation}
\label{eq:box_alpha}
\langle {\rm grad}\varphi, {\rm grad} \varphi \rangle
={\Box \alpha \over \alpha}.
\end{equation}
This means that, given the vector field $\psi$ with covariant components
(i.e., its $1$-form realization)
\begin{equation}
\label{eq:phasefun}
\psi_{\mu}=\alpha^{2} \nabla_{\mu} \varphi,
\end{equation}
one can first look for solutions of the divergenceless condition
\begin{equation}
\label{div_psi}
{\rm div}(\psi)=\nabla^\mu \psi_\mu=0.
\end{equation}
As a second step, provided that the form obtained for $\psi$ satisfies the integrability condition
($\psi$ having to be hypersurface orthogonal, i.e., curl-free in order to be a gradient) 
\begin{equation}
\label{hyp_orthog}
\psi_{[\mu} \; \nabla_{\nu} \; \psi_{\kappa ]}=0\,,
\end{equation}
one obtains from Eq. \eqref{eq:box_alpha} the equation
\begin{equation}
\label{eq:nonlineare}
\alpha^{3}\Box \alpha=\langle \psi, \psi \rangle ,
\end{equation}
to be solved for $\alpha$ once a solution for $\psi$ is already obtained. 
Third, Eq. \eqref{eq:phasefun} yields the phase function 
$\varphi$ by solving the first-order equations
\begin{equation}
\label{phase_fun}
\nabla_{\mu}\varphi=\alpha^{-2} \psi_{\mu}.
\end{equation}
Of course, the reader may wonder whether it is really a good idea to turn
a linear second-order hyperbolic equation into a non-linear problem.
A possible answer is that not only does the amplitude and phase language provide 
the appropriate tool for studying the parametrix (i.e., an approximate 
Green function) of a hyperbolic operator, but in Kasner space-time the work
of Ref. \cite{EM} has proved that, if one can find a pair of auxiliary
$1$-forms ($\psi,\rho$), $\psi$ being real-valued and divergenceless as above, 
$\rho$ being complex-valued and fulfilling
\begin{equation}
{\rm div}(\rho)+\langle \rho, \rho \rangle =0\,,
\end{equation}
it is then possible to evaluate the amplitude $\alpha$ by 
solving the first-order non-linear equation
\begin{equation}
\nabla_{\gamma}(\log \alpha)+i {\psi_{\gamma}\over \alpha^{2}}
=\rho_{\gamma}.
\end{equation}
One then obtains a promising framework for studying local
as well as weak solutions of wave equations by means of a
system of first-order equations, where the equation for $\rho$
is non-linear. The reader can find the explicit expressions of the
desired $1$-forms, amplitude and phase for Kasner in Ref. 
\cite{EM}, which has been also a motivation for the present study.
Another interesting possibility is to consider the two equations 
\eqref{eq:box_alpha} and \eqref{div_psi} as a new set of \lq\lq model 
equations," to be studied independently of their original relation with   
the massless scalar field wave equation: the novelty is indeed the 
evolution of a scalar field interacting with a given gravitational 
background and sourced by itself in a proper manner.
This second point of view turns this problem into a new one, which is worth 
discussing in terms of applications which can better clarify its usefulness.

Hereafter, we restrict ourselves to the use of $\psi$ only and,
in order to obtain a physics-oriented scheme, we find it convenient to
re-express its covariant components in the form
\begin{equation}
\psi_{\mu}=f_{c}J_{\mu},
\end{equation}
where $f_{c}$ is a freely specifiable coupling constant, and $J_{\mu}$ 
is a current covector (which we do not consider here as restricted 
by any causality condition). 
The non-linear equation \eqref{eq:nonlineare} bears a clear resemblance 
with the Ermakov-Pinney non-linear ordinary differential equation \cite{E,P}
\begin{equation}
y''(x)+p(x)y(x)=q(x)y^{-3}(x),
\end{equation}
upon setting $p=0$ therein, which is why we refer to it as the Ermakov-Pinney equation
in curved space-time\footnote{Note that a harmonic time dependence leads to a
term linear in $y$ and hence corresponds to the $p \not =0$ case.}.
More precisely, by the Ermakov-Pinney-like equation in a generic 
curved space-time we mean hereafter the coupled set of equations
\begin{equation}
\label{eq:=EP}
\alpha^3 \Box \alpha =f_{c}^{2}  ||J||^{2}, \; \qquad  \nabla_\mu J^\mu=0 ,
\end{equation}
where $||J||^2 =\langle J, J \rangle  
=\epsilon   |J^\mu J_\mu|$, with $\epsilon=-1,0,1$ for 
$J$ timelike, null, spacelike, respectively. Since for $\epsilon=0$ one finds for $\alpha$ 
just the scalar wave equation \eqref{eq:1} we started from, 
we shall limit ourselves to studying timelike and spacelike currents.
Our $\alpha$ can be seen as a real, self-interacting scalar field, gravitationally 
interacting with the background space-time\footnote{Generalizations with $\alpha$  
a complex scalar field are also possible.} and sourced by a divergence-free vectorial current 
$J^\alpha$. The latter is consistently obtained by solving the divergence equation in the 
assigned background. The equation for $\alpha$ can be also cast in the form
\begin{equation}
\label{eq:EP2}
\alpha^3 \Box \alpha= f_{c}^{2} J^\mu J_\mu \equiv 
\epsilon f_{c}^{2} |J^\mu J_\mu| .
\end{equation}
In the spacelike and timelike cases non-linearities come 
into play, leading to new interesting features, as will be shown in the
following sections. We point out that the divergence-free condition for $J$
does not pose any limitation to $f_c$ since this is a factorizable constant. 
Without any loss of generality we may adapt it in the $\Box\alpha$ equation 
so that it is always dimensionless. 

Some formal simplification or extensions can still be considered. For example,
it can be convenient to separate magnitude and direction of $J$, 
i.e., $J^\mu = \rho u^\mu$, $u\cdot u=\epsilon$, so that
\begin{equation}
u^{\mu}\nabla_{\mu}\rho+\rho \nabla_{\mu}u^{\mu}=0.
\end{equation}
Equivalently,
\begin{equation}
\nabla_u \rho +\rho \Theta(u)=0\,,\qquad \Theta(u)=\nabla_\mu u^\mu .
\end{equation}
The scalar $\Theta(u)$, for example, has the geometrical meaning of  
the expansion of the congruence $u$ if $u$ is timelike.

Thus, with the introduction of $\rho$, 
the equation for $\alpha$, Eq. \eqref{eq:EP2}, reads as
\begin{equation}
\label{eq:EP3}
\alpha^3 \Box \alpha= \epsilon f_{c}^{2} \rho^{2} .
\end{equation}
One can also study the geometrical content of the condition 
\eqref{hyp_orthog} in terms of kinematical properties of the congruence $u$.
In fact \eqref{hyp_orthog} can be cast in the equivalent form
\begin{equation}
\label{hyp_orthog2}
\eta_{\lambda\mu\nu\sigma} (\rho u^\mu)\nabla^\nu (\rho u^\sigma)=0 ,
\end{equation}
where $\eta_{\lambda\mu\nu\sigma}$ is the unit volume $4$-form 
(with the defining property $\eta_{\hat 0\hat 1\hat 2\hat 3}=1$ 
in an orthonormal frame $e_{\hat \alpha}$).
Expanding the covariant derivative Eq. \eqref{hyp_orthog2} becomes
\begin{equation}
\eta(u)_{\lambda\mu\nu}\nabla^\mu u^\nu =0,
\end{equation}
having introduced the notation 
$\eta(u)_{\lambda\mu\nu}=u^\sigma \eta_{\sigma\lambda\mu\nu}$. 
The $3$-form $\eta(u)_{\lambda\mu\nu}$ is orthogonal to $u$, 
i.e., any contraction by $u$ vanishes identically. 
In the timelike case ($u\cdot u=-1$) the above condition implies 
that $u$ defines a vorticity-free congruence of world lines (analogous 
properties hold also in the case of a null and a spacelike congruence).

We notice that in view of other extensions, it is possible to consider the counterpart 
of the above Eq. \eqref{eq:EP3} in a gauge-theory context, where the full
covariant derivatives receive a contribution from a 
gauge potential $A^{\beta}{}_{\mu}(x)$,
i.e. a Lie-algebra-valued $1$-form, according to\footnote{We follow the
convention according to which Greek indices from the beginning of the 
alphabet are Lie-algebra indices. When necessary, this index specification is 
explicitly repeated in the text to avoid confusion.}
\begin{equation}
\label{gauge_eq}
\partial_\mu \to \nabla_\mu + G_\beta A^{\beta}{}_{\mu}(x) 
\end{equation}
with $G_\beta$ the generators of the symmetry group.
In the case of a scalar field Eq. \eqref{gauge_eq} becomes
\begin{equation}
\partial_\mu \to D_\mu=\nabla_\mu + q A_{\mu}(x) ,
\end{equation}
implying
\begin{equation}
D_\mu D^\mu \alpha =\Box \alpha + q \alpha \nabla_\mu A^\mu 
+2q A^\mu \nabla_\mu \alpha+q^{2}A_{\mu}A^{\mu}\alpha .
\end{equation}

On reverting now to the central aim of our paper, a naturally occurring question is whether such 
a scheme is equally successful as the other direct (standard) approach.
Properly speaking, the two methods are equivalent, even if in one case one aims at solving a single, 
linear equation, Eq. \eqref{eq:1}, whereas in the other case the equations to be solved are two, 
Eqs. \eqref{eq:box_alpha} and \eqref{div_psi}, coupled and with one of the two (Eq. \eqref{eq:box_alpha}) 
which is also non-linear. Recently, the work in Refs. \cite{EBD,EM}, following the second approach   
in a Kasner spacetime, has obtained the exact expression
for amplitude $\alpha$ and phase function $\varphi$ in the integral representation of
the solution for given initial conditions.
The success achieved in the simple Kasner context has (mainly) exploited the fact 
that one of the two equations, the divergence equation, was particularly easy to solve.
This solution, in turn, has been \lq\lq driving" in a sense the corresponding solution of the 
second equation as well. A naturally occurring question is therefore whether this approach can be equally 
successful also in other cases. In order to answer this question we have 
analyzed, in Secs. 2-4 below, three typical exact solutions: 
Schwarzschild, de Sitter, gravitational wave, solving in all cases the Ermakov-Pinney equation.
Later on we solve for the phase function that obeys Eq. \eqref{phase_fun}, 
and hence we plot the solutions $\alpha \cos \varphi$ of the
wave equation \eqref{eq:1}. Concluding remarks are made in Sect. 5. 

\section{Explicit examples}

We will discuss hereafter the cases of Schwarzschild, de Sitter and a single plane gravitational 
wave space-times. Their examination in specific contexts will contribute 
to clarify the physical content of the Ermakov-Pinney equation.
We will solve the coupled set of equations \eqref{eq:EP2} and \eqref{div_psi}
\begin{equation}
\label{eq:EPall}
\alpha^3 \Box \alpha= 
f_{c}^{2}  J^\mu J_\mu  \,,\qquad \nabla_\mu J^\mu=0 ,
\end{equation}
also providing (analytically or numerically) the phase function $\varphi$ 
defined by Eq. \eqref{phase_fun} above
\begin{equation}
\nabla_{\mu}\varphi=\alpha^{-2}f_{c}J_{\mu} ,
\end{equation}
as well as the product $X={\rm Re}(\chi)=\alpha \cos \varphi$, 
which is the real part of the desired solution of Eq. \eqref{eq:1}.

We are well aware of existing literature concerning exact or approximate 
solutions of the massless scalar field equation in these space-times, 
also in the case of non-vanishing source terms. For example, in recent 
years the gravitational self-force approach \cite{Detweiler:2002gi} 
has analyzed sources (either the massless scalar field equation or 
the full set of gravitational perturbations) consisting of a  
massive particle with an energy-momentum tensor having support only 
along a world line (i.e., Dirac-delta singular along that world line). 
It is clear that these results can be translated into our Ermakov-Pinney-like 
equation approach. However, the main point of our work is to examine the 
converse: to find a solution of a non-linear, massless scalar field equation 
which is also sourced by itself, like in the case of self-interacting fields, 
instead of the problem of identifying the perturbations induced on a 
given background by a known source. Due to non-linearities existing here this 
is a non-trivial problem, which we have introduced \lq\lq preliminarly" 
(because it is easy to think of follow-up papers) by providing and discussing 
several, explicit and simple examples in some familiar space-times. Indeed, 
one could have started directly from an Ermakov-Pinney-like equation in a 
given curved background, ignoring  the initial derivation, i.e., its relation 
with the massless scalar field equation. Summarizing, the Ermakov-Pinney-like  
equation is a model equation concerning a massless scalar field in interaction 
with a gravitational background and (non-linearly) with itself. Existence 
of solutions and the associated physical meaning are all features under investigation.

\subsection{Schwarzschild space-time}

Let us consider the case of a Schwarzschild space-time with metric written in standard 
Schwarzschild coordinates $x^\mu=(t,r,\theta,\phi)$
\begin{equation}
\label{sch_met}
ds^2=-\left(1-\frac{2M}{r} \right) dt^2 +\frac{dr^2}{\left(1-\frac{2M}{r} \right)}
+r^2 (d\theta^2+\sin^2\theta d\phi^2).
\end{equation}
Looking for particular solutions instead of general formulae, 
one can provide interesting examples, as shown below.

\begin{enumerate}
  \item $J$ timelike
  
Upon assuming $J$ timelike ($\epsilon=-1$) and aligned with the time lines, a simple solution 
of the divergence equation ${\rm div} (J)=0$ is given by the vector field\footnote{In general,
all our formulae for currents are particular cases of the general expressions
$$
J=J^{\mu}{\partial \over \partial x^{\mu}}, \;
J_{\nu}=J^{\mu}g_{\mu \nu}, \;
J^{\flat}=J_{\nu}dx^{\nu}.
$$}
\begin{equation}
J= \frac {A^{2}}{(1- \frac {2M}{r})}{\partial \over \partial t} 
\Longrightarrow 
J^{\flat}=-A^{2}dt ,
\end{equation}
where $A=A(r)$. Thus, the squared pseudo-norm of $J$ is equal to
\begin{equation}
\langle J, J \rangle =- \frac {A^{4}}{(1-\frac{2 M}{r})}.
\end{equation}
On setting $f_{c}^{2}\to f_{c}^{2} M^{-2}$ (for dimensional reasons)
and $\alpha=A(r)$, Eq. \eqref{eq:EP2} can be rewritten in the form
\begin{equation}
\label{EP_new}
A^3 \Box A=\frac {f_{c}^{2}}{M^2} \langle J,J \rangle 
=-\frac {f_{c}^{2}}{M^2} \frac {A^{4}}{(1-\frac{2 M}{r})} .
\end{equation}
Thus, the Ermakov-Pinney equation becomes
\begin{equation}
\label{eq_alpha_gc}
\frac{d^2A}{dr^2}=\frac {2(M-r)}{r^{2}(1-\frac{2 M}{r})}\frac{dA}{dr} 
- \frac{f_{c}^{2}}{M^{2}(1-\frac{2 M}{r})^{2}}A .
\end{equation}
This form of the equation suggests defining the independent
variable $x$ via
\begin{equation}
\label{rho_def}
\varrho \equiv \frac {r}{M}\,,\qquad 
x \equiv \frac {\varrho}{2}-1 , 
\end{equation}
such that $x\in [0,\infty[$. 
Hence we obtain the linear second-order equation in the variable $\varrho$
\begin{equation}
\label{eq_alpha_gc_in_rho}
\frac{d^2A}{d\varrho^2}=\frac {2(1-\varrho)}{\varrho(\varrho-2)}\frac{dA}{d\varrho} 
- \frac{f_{c}^{2}\varrho^2}{ (\varrho-2)^{2}}A\,,
\end{equation}
as well as in the variable $x$
\begin{equation}
\label{eq_A_x_timelike}
\frac{d^2A}{dx^2}+ \frac{(1+2x)}{x(x+1)}\frac{dA}{dx} 
-\xi^2 \frac{(x+1)^{2}}{x^{2}}A=0 ,
\end{equation}
where $\xi=2if_c$. Note that this equation contains the parameter 
$f_c$ only through the combination $\xi^2$, 
and hence is invariant under the map 
$\xi \to -\xi=\bar \xi$. A simple consequence of this is that, 
given a solution $A_1(\xi,x)$, a second, independent 
solution is $A_2(\xi,x)=A_1(-\xi,x)$.
Indeed, on denoting by ${\rm HeunC}$ the confluent Heun function
\cite{Heun}, the general solution of such an equation is
\begin{eqnarray}
A(\xi,x)&=&C_1 A_1(\xi,x)+C_2A_2(\xi,x)\nonumber\\
&\equiv &e^{-\xi x}\left[C_{1}x^{\xi}H_{1}(\xi, x)
+C_{2}x^{-\xi}H_{2}(\xi, x)\right],
\end{eqnarray}
where $C_{1}$ and $C_{2}$ are arbitrary constants, and
we have defined
\begin{eqnarray}
H_{1}(\xi, x) &\equiv& {\rm HeunC}\Bigr[2\xi ,2\xi ,0,
2\xi^2,-2\xi^2,-x \Bigr],
\nonumber\\
H_{2}(\xi, x) &\equiv& {\rm HeunC} \Bigr[2\xi ,-2\xi,0,
2\xi^2,-2\xi^2,-x \Bigr].
\end{eqnarray}
One can then prove the above mentioned \lq\lq doubling" property, $A_2(\xi,x)=A_1(-\xi,x)$, 
by virtue of the identity
\begin{equation}
H_{1}(-\xi, x)=e^{-2\xi x} H_{2}(\xi, x) .
\end{equation}

\begin{figure}
\includegraphics[scale=0.35]{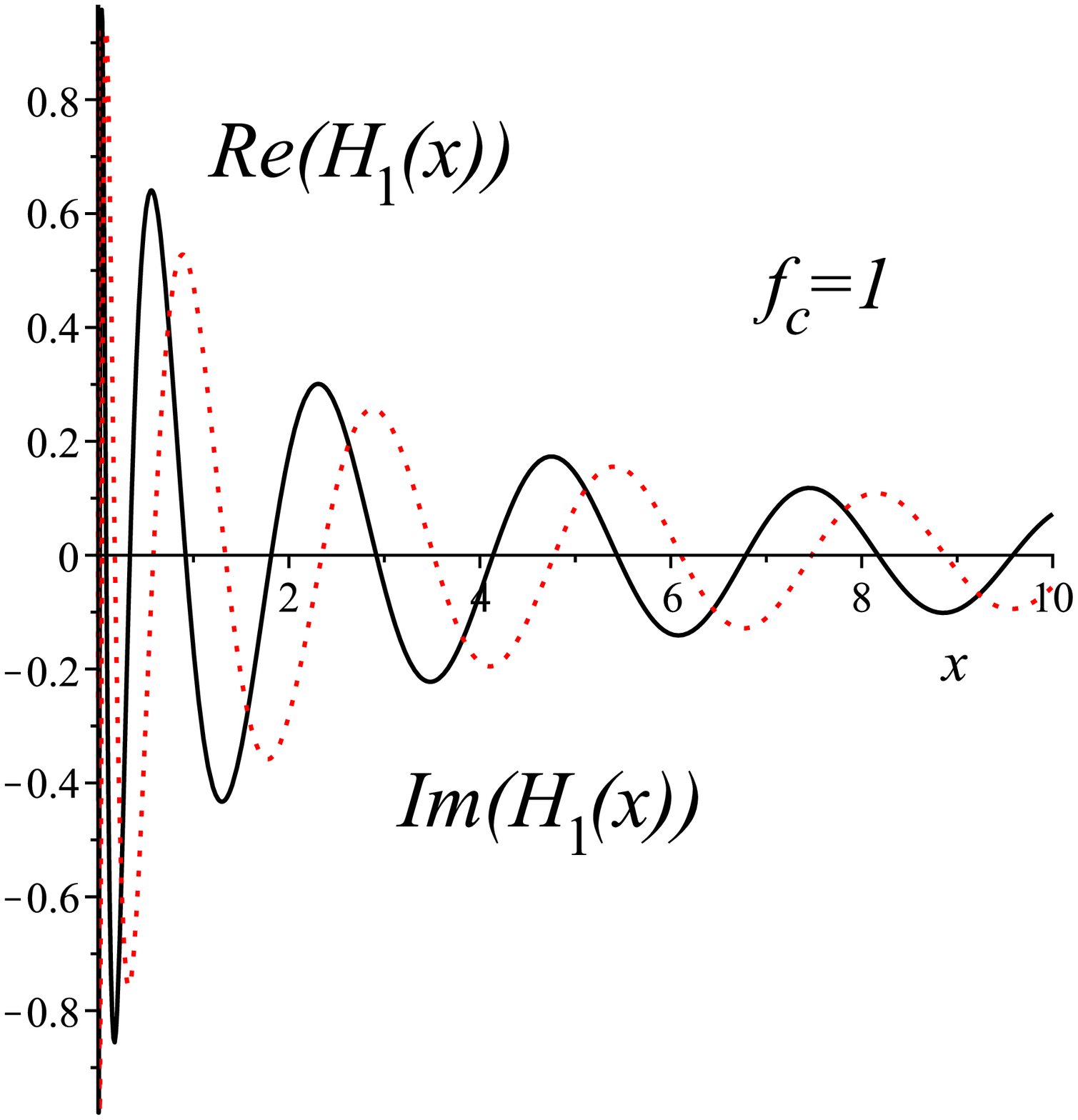}\\
\includegraphics[scale=0.35]{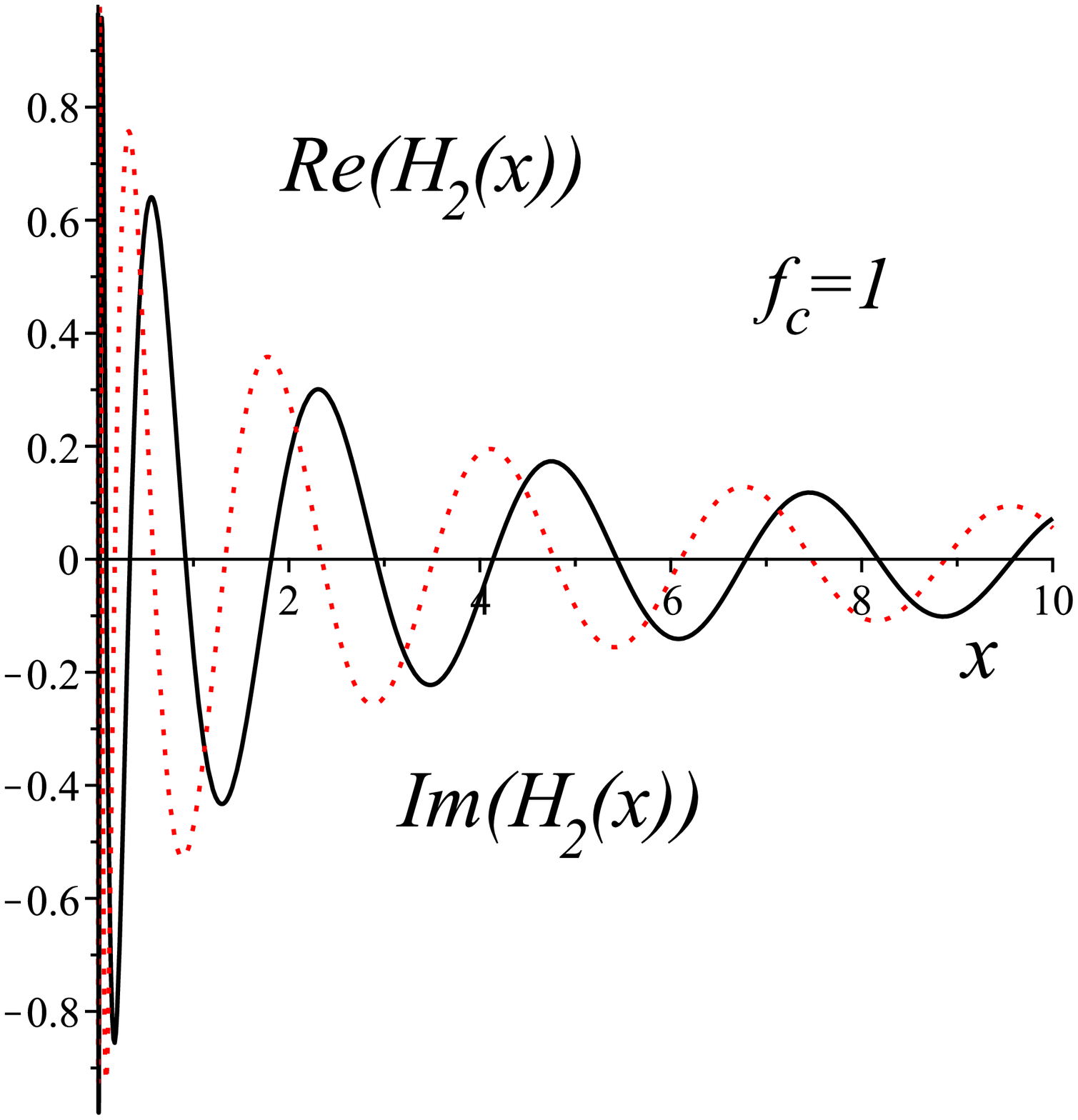}  
\caption{\label{fig:1}Schwarzschild solution and a timelike current 
vector: The real (solid curve, black online) 
and imaginary (dotted curve, red online) parts of the functions $H_1(x)$ 
and $H_2(x)$ are shown in the case $f_c=1$.}
\end{figure}

The dependence on the parameter $f_c$ in the solutions of Eq. \eqref{eq_alpha_gc} can be 
studied by integrating numerically the equation. The case $f_c>0$ of interest 
here corresponds to spatially damped oscillations, as stated above. 
The situation is illustrated in Fig. \ref{fig:2}.
In this case the phase function reduces to
\begin{equation}
\varphi=-f_c t +\varphi_0 ,
\end{equation}
where $\varphi_0$ is an integration constant, and then the associated 
quantity $X=\alpha \cos \varphi$ becomes 
$X=A(r)\cos (f_c t -\varphi_0)$.
We conclude this sub-section by answering the following two questions: 
\begin{enumerate}
  \item Can we obtain a simple understanding of the solutions near the horizon and at infinity?
  \item Are there regular solutions at the horizon?
\end{enumerate}

In order to answer the first question it is worth mentioning 
that the equation for $A$ can be always reduced to its normal form. 
For example, the rescaling
\begin{equation}
A(\varrho) =\frac{{\mathbb A}(\varrho)}{\sqrt{\varrho (\varrho-2)}} 
\end{equation}
implies for Eq. \eqref{eq_alpha_gc_in_rho} the following form:
\begin{equation}
\frac{d^2 {\mathbb A}}{d\varrho^2}+V(\varrho){\mathbb A}=0 ,
\end{equation}
where the \lq\lq potential" $V(\varrho)$ is given by
\begin{equation}
V(\varrho)=\frac{f_c^2\varrho^4+1}{\varrho^2 (\varrho-2)^2} .
\end{equation}
In the limit $\varrho \to 2$ we find
\begin{equation}
V(\varrho)\approx \frac{16 f_c^2+1}{4(\varrho-2)^2}
\end{equation}
so that
\begin{equation}
{\mathbb A}(\varrho)\approx  \sqrt{\varrho-2} \left[ 
C_1 \sin(2 f_c\ln(\varrho-2))+ C_2 \cos(2 f_c\ln(\varrho-2))\right],
\end{equation}
while in the limit $\rho\to \infty$
\begin{equation}
V(\varrho)\approx  f_c^2 ,
\end{equation}
so that
\begin{equation}
{\mathbb A}(\varrho)\approx C_1 \sin(f_c \varrho)+C_2 \cos(f_c \varrho).
\end{equation}
Recombining then these results in $A(\varrho)$ one recovers the already 
found behaviours at the horizon and at infinity.

In order to answer the second question we first of all note that a 
\lq\lq horizon-penetrating" coordinate system will be more appropriate
for such an analysis. However, one can try to solve the equation for 
$A$, e.g.,  Eq. \eqref{eq_alpha_gc_in_rho}, by series. Formally one 
cannot look for solutions of the type
\begin{equation}
A(\varrho)=\sum_{k=0}^\infty c_k (\rho-2)^k ,
\end{equation}
because of the singularity of the equation at $\rho=2$ (of higher order 
than for $\rho=0$, for example). It is easy instead to find solutions 
which are \lq\lq regular at the origin," $\rho=0$ 
like series solutions of the form
\begin{eqnarray}
\label{sol_series}
A(\varrho)&=&1 -\frac{1}{64} f_c^2\varrho^4 -\frac{9}{800} f_c^2\varrho^5 
-\frac{37}{5760} f_c^2 \varrho^6 \nonumber\\
&&-\frac{319}{94080} f_c^2 \varrho^7+\left(-\frac{743}{430080} f_c^2
+\frac{1}{16384} f_c^4\right)\varrho^8\nonumber\\
&&+\left(-\frac{2509}{2903040} f_c^2+\frac{713}{8294400} f_c^4\right)\varrho^9
+O(\varrho^{10}) . 
\end{eqnarray}
By going to very high orders in the $\varrho$ expansion (an easy task for 
nowadays computers) one can then extend this solution beyond the horizon.
Another route would be that of looking for a Post-Newtonian-like solution, 
i.e., to expand the equation in powers of $M$ 
(i.e., of the gravitational radius $GM/c^2$)
and look for perturbative solutions, also expanded in series of $1/c$.
The result is that one easily identifies two kinds of solutions
\begin{eqnarray}
A_1(r) &=& 1-\frac{1}{6}\eta^4 f_c^2 \frac{r^2}{M^2}-\frac{5}{3}\eta^6 
f_c^2 \frac{r}{M}  +O(\eta^8)  , \nonumber\\
A_2(r)&=& \frac{M}{r}+\eta^2\frac{M^2}{r^2}+\eta^4 \left(-\frac12  
f_c^2\frac{r}{M}+\frac{4M^3}{3 r^3}  \right)\nonumber\\
&&+\eta^6 \left(\frac{2M^4}{r^4} -4f_c^2\ln\left(\frac{r}{R} 
\right)\right)  +O(\eta^8) ,
\end{eqnarray}
where we have denoted by $\eta=1/c$ a place-holder in the Post-Newtonian 
expansion, while $R$ is an integration constant with the dimensions of a length. 
It is easy to see that the first solution $A_1(r) $ is \lq\lq purely 
ingoing," i.e., starts regular at the origin while the second one, 
$A_2(r)$ is \lq\lq purely outgoing,"i.e., starts regularly at infinity.
As soon as the Post-Newtonian order (i.e., $\eta$) increases, deviations 
from the mentioned regular behaviour start anyway (for example in 
$A_2(r)$ there appear logarithms).
In this perturbative case having regularity in $r=0$ for $A_1(r)$ and not 
at the horizon is understandable: the expansion in $M$ implies that in the 
equation the horizon will be built by adding sufficiently many orders. 
In other words, the expanded equation \lq\lq does not know" the horizon, 
until this is practically rebuilt after summing many orders. 

These two families of independent solutions can be matched together to 
recover the solution \eqref{sol_series}, or to find a combination with proper 
regularity conditions, for example at the horizon (always built in 
post-Newtonian sense, i.e., order-by-order), but this is not an easy task 
and we will not insist anymore on this point. \\

Eventually, the above discussion can be equally well performed 
in most of the subsequent situations. We will avoid this repetition.

\begin{figure}
\includegraphics[scale=0.35]{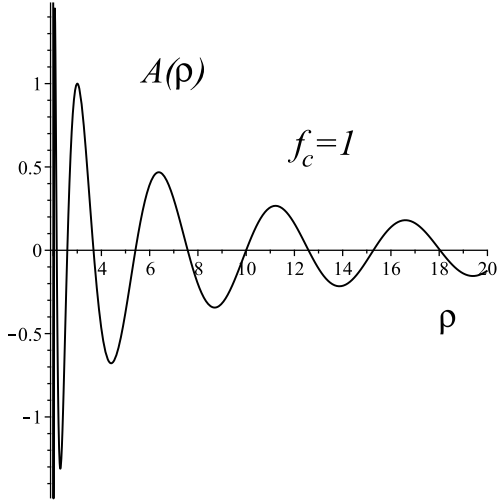} 
\caption{\label{fig:2} Schwarzschild solution and a timelike current vector: 
The full solution  $A(\varrho)$ is plotted in the case $f_c=1$, as an example.
Initial conditions are chosen so that $A(3)=1$ and $A'(3)=0$. 
The oscillating and spatially damped  
behaviour is further enhanced as soon as the value of $f_c$ increases.}
\end{figure}

\item $J$ spacelike
  
For the case of spacelike current, we consider  
\begin{equation}
J= \frac{M^2 \cos^{2} \omega t}{r^{2}}{\partial \over \partial r}
\Longrightarrow
J^{\flat}=\frac {M^2\cos^{2}\omega t}{r^{2}\left(1-\frac {2 M}{r}\right)}dr,
\end{equation}
with
\begin{equation}
\langle J,J \rangle =\frac{M^4 \cos^4 \omega t}{r^4 \left(1-\frac {2 M}{r}\right)}.
\end{equation}
This current satisfies the divergenceless condition, although it is not 
the most general form of current that satisfies such a property.
Now we look for the amplitude function $\alpha$ in the factorized form
\begin{equation}
\alpha(r,t)=A(r)\cos \omega t .
\end{equation}
The Ermakov-Pinney equation reads 
\begin{equation}
\alpha^{3} \Box \alpha=\frac{f_{c}^{2}}{M^{2}} \langle J, J \rangle ,
\end{equation}
leading to the following non-linear equation for $A(r)$:
\begin{equation}
\label{eq_A_J_spat_schw}
\frac{d^2A }{dr^2}= \frac {2(M-r)}{r (r-2 M)}\frac{dA}{dr} 
+ \frac {1}{(r-2M)^{2}} 
\left[ \frac {f_{c}^{2}M^{2}}{r^{2}A^{3}}-r^{2}\omega^{2}A \right],
\end{equation}
that we have solved numerically.
On passing to the variable $\varrho$, defined in Eq. \eqref{rho_def}, 
Eq. \eqref{eq_A_J_spat_schw} becomes
\begin{equation}
\frac{d^2A}{d\varrho^2} =-2 \frac {(\varrho-1)}{\varrho(\varrho-2)}\frac{dA}{d\varrho} 
- \frac {(\varrho^{2}A^2 \Omega-f_{c})(\varrho^{2}A^2 \Omega 
+ f_{c})}{\varrho^{2}(\varrho-2)^{2}A^{3}(\varrho)},
\end{equation}
where $\Omega=M \omega$ is dimensionless.

Next, we look for the phase function $\varphi$ that solves the
first-order equation
\begin{equation}
{\partial \varphi \over \partial r}
=\alpha^{-2} f_{c}J_{r},
\end{equation}
and is therefore found to be
\begin{equation}
\varphi=f_c M^2 \int^r  \frac{ dr}{A(r)r^{2}(1-\frac {2 M}{r})}  +
\varphi_0.
\end{equation}

\begin{figure}
\includegraphics[scale=0.35]{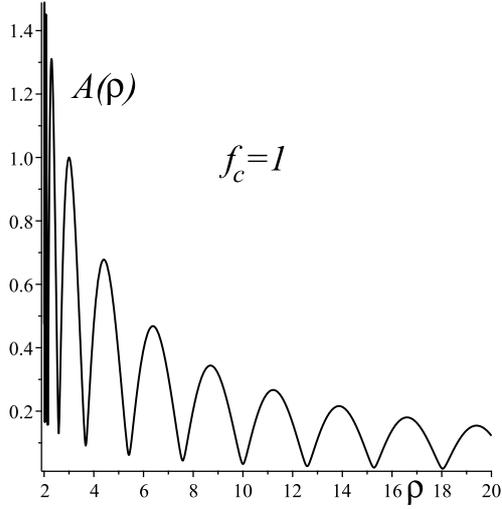}
\caption{\label{fig:3} Schwarzschild solution and a spacelike current vector: The full solution    
$A(\varrho)$ is plotted in the case $f_c=1$ and $\Omega=1$ (black online), as an example.
The initial conditions are chosen so that $A(3)=1$ and $A'(3)=0$.}
\end{figure}

The case $f_c>0$ of interest here corresponds to damped 
oscillations, as in the timelike case. The difference is that now the function 
remains positive while being subject to damping. 
The larger the value of $f_c$, the more frequent are 
the oscillations. The behaviour of $A$ is illustrated in Fig. \ref{fig:3} for 
$f_{c}=1$, whereas Fig. \ref{fig:4} shows the behaviour of the phase $\varphi$ as a 
function of $\varrho$, again in the case $f_c=1$ and $\Omega=1$.

\begin{figure}
\includegraphics[scale=0.35]{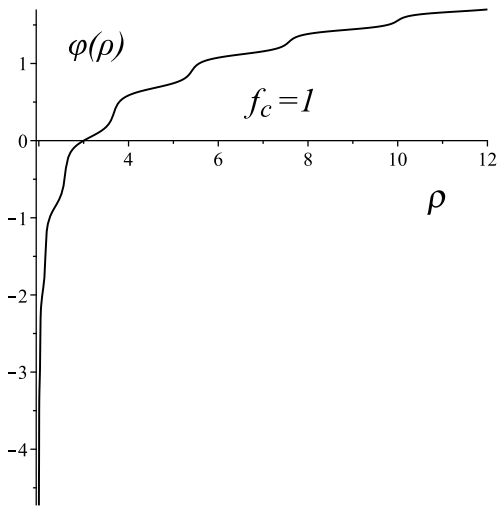}
\caption{\label{fig:3bis} Schwarzschild solution and a spacelike current vector: The phase    
$\varphi(\varrho)$ is plotted in the case $f_c=1$ and $\Omega=1$ (black online), as an example.
The initial conditions are chosen so that $\varphi(3)=0$, besides 
$A(3)=1$ and $A'(3)=1$ (as above in Fig. \ref{fig:3}).}
\end{figure}

\end{enumerate}

\section{de Sitter space-time} 

The de Sitter space-time metric in spherical-like coordinates $x^\mu=(t,r,\theta,\phi)$ 
has a squared line element
\begin{equation}
\label{ds_met}
ds^2=-(1-H^2r^2) dt^2 +\frac{dr^2}{(1-H^{2}r^{2})}
+r^2 (d\theta^2+\sin^2\theta d\phi^2),
\end{equation}
and it is formally close to the Schwarzschild metric, except for a different expression 
for the $g_{tt}=-1/g_{rr}$ metric component and the presence of a cosmological 
horizon at $r_h=1/H$, so that in this case $r\in[0,1/H]$. 
This metric is a solution of Einstein's equations with a cosmological 
constant $\Lambda=3H^2$ \cite{S2003}.

\begin{enumerate}
  
\item $J$ timelike
  
We consider a divergenceless current having the vector field description
\begin{equation}
J= \frac{A^{2}(r)}{(1-H^{2}r^{2})}{\partial \over \partial t}
\Longrightarrow
J^{\flat}=-A^{2}(r)dt ,
\end{equation}
with
\begin{equation}
\langle J,J\rangle = -\frac{A^{4}(r)}{(1-H^{2}r^{2})} .
\end{equation}

The Ermakov-Pinney equation for $\alpha=A(r)$ is
\begin{equation}
A^{3}\Box A=f_{c}^{2}H^{2}\langle J,J \rangle ,
\end{equation}
and leads, upon defining $\rho \equiv r H \in (0,1)$,
to the linear equation
\begin{equation}
\frac{d^2A}{d\varrho^2}= \frac {2(2\varrho^{2}-1)}{\varrho(1-\varrho^{2})}\frac{dA}{d\varrho} 
- \frac {f_{c}^{2}}{(\varrho^{2}-1)^{2}}A .
\end{equation} 
This equation can be solved explicitly in the form
\begin{equation}
A(\varrho)=C_1{\mathcal A}(f_c,\varrho)+C_2 {\mathcal A}(-f_c,\varrho) ,
\end{equation}
where 
\begin{equation}
{\mathcal A}(f_c,\varrho) \equiv {1 \over \rho} 
\left( \frac {\varrho-1}{\varrho+1} \right)^{{i \over 2}f_{c}}
(\varrho+i f_{c}),
\end{equation}
and $C_1$ and $C_2$ are integration 
constants.\footnote{Note that, as discussed above in the 
Schwarzschild case, the second term here is obtained from the first by replacing $f_c\to-f_c$.}
Moreover, the phase function $\varphi$ solves the equation
\begin{equation}
{\partial \varphi \over \partial t}=\alpha^{-2}f_{c}J_{t}=-f_{c},
\end{equation}
which therefore yields
\begin{equation}
\varphi=-f_{c}t+\varphi_0,
\end{equation}
where $\varphi_0$ is an additive constant.

The case $f_c>0$ of interest here corresponds to the \lq\lq deformed bell" behaviour 
shown in Fig. \ref{fig:5a}. Since both the phase and associated $X$ variables are simply 
related to $A(\varrho)$ we will not display their plots.

\begin{figure}
\includegraphics[scale=0.35]{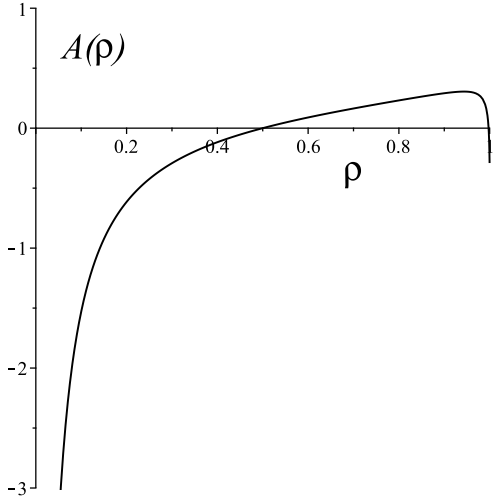} 
\caption{\label{fig:5a} de Sitter solution and a timelike current vector: The solution  
$A(\varrho)$ is plotted in the case $f_c=1$ (black online) and initial conditions 
$A(0.5) = 0$, $A'(0.5)=1$.} 
\end{figure}

\item $J$ spacelike
  
We here assume a divergenceless current vector field in de Sitter reading as
\begin{equation}
J= \frac {\cos^{2}\omega t}{H^2 r^{2}}{\partial \over \partial r}
\Longrightarrow 
J^{\flat}= \frac {\cos^{2}\omega t}{H^2 r^{2}(1-H^{2}r^{2})}dr.
\end{equation}
By using the variable $\rho \equiv r H$ and (dimensionless) parameter 
$\Omega \equiv {\omega \over H}$, the equation for 
$\alpha(t,r)=A(r)\cos\omega t$ becomes the following non-linear equation for $A$:
\begin{equation}
\frac{d^2A}{d\varrho^2}=2 \frac {(2\varrho^{2}-1)}{(1-\varrho^{2})\varrho}\frac{dA}{d\varrho}
- \frac {(\Omega A^{2} \varrho^{2}-f_{c})(\Omega A^{2}\varrho^{2}+f_{c})}
{(1-\varrho^{2})^{2}\varrho^{4}A^{3}},
\end{equation}
that we have solved numerically.
The phase $\varphi$ is then such that
\begin{equation}
\varphi=\frac{f_c}{H^2}\int^r \frac{dr}{A^{2}r^2(1-H^2r^2)}  +\varphi_0.
\end{equation}

The behaviour of $A(\varrho)$ is shown in Fig. \ref{fig:4}, whereas that of the 
phase $\varphi(\rho)$ in Fig. \ref{fig:ds5}.
\begin{figure}
\includegraphics[scale=0.35]{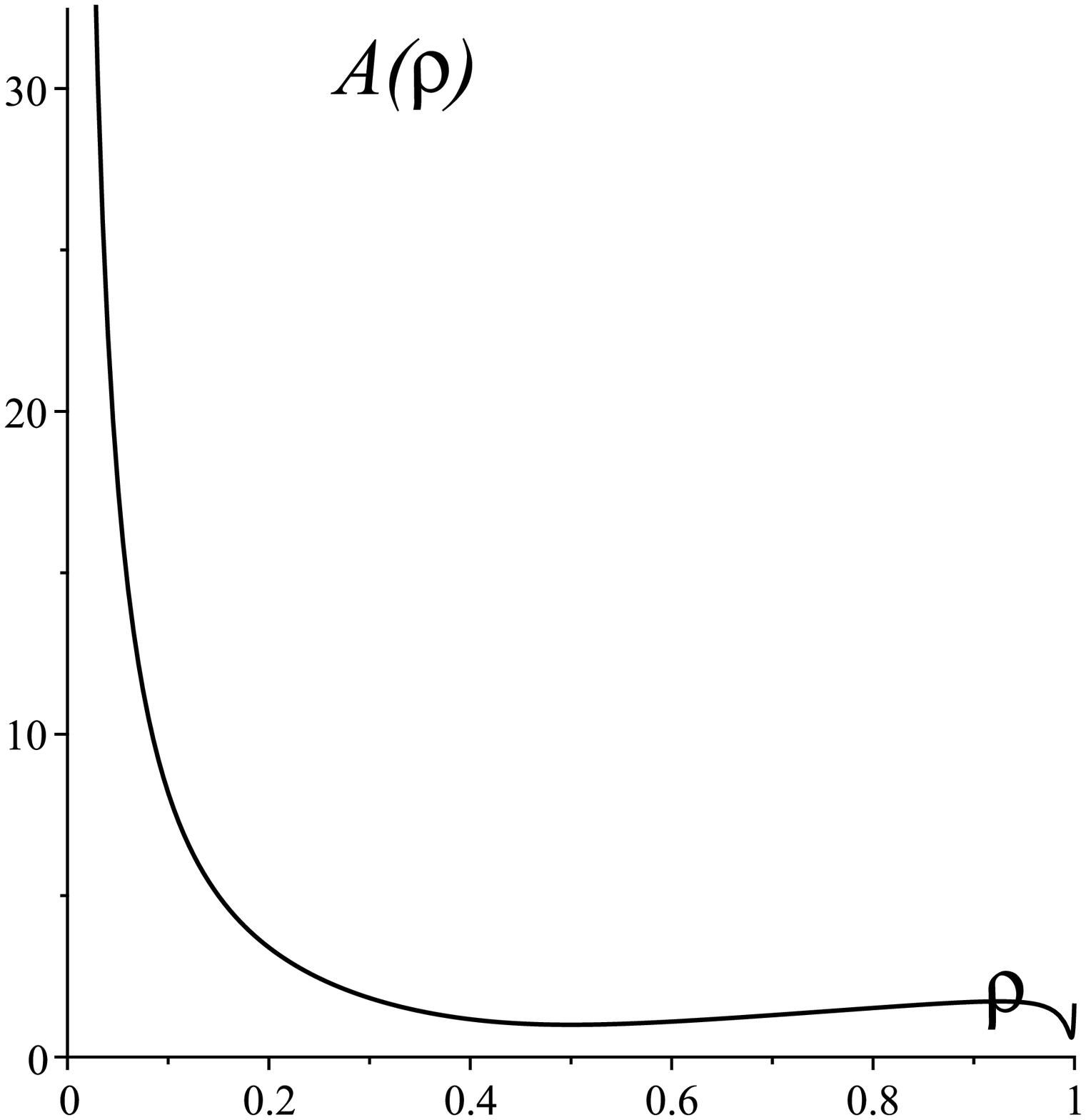}
\caption{\label{fig:4} de Sitter solution and a spacelike current vector:  
The solution $A(\rho)$ is plotted in the case $f_c=1$ (black online).
The parameters are fixed as $\Omega=1=H$ and the initial conditions are chosen 
so that $A(0.5)=1$ and $A'(0.5)=0$, as an example.}
\end{figure}

\begin{figure}
\includegraphics[scale=0.35]{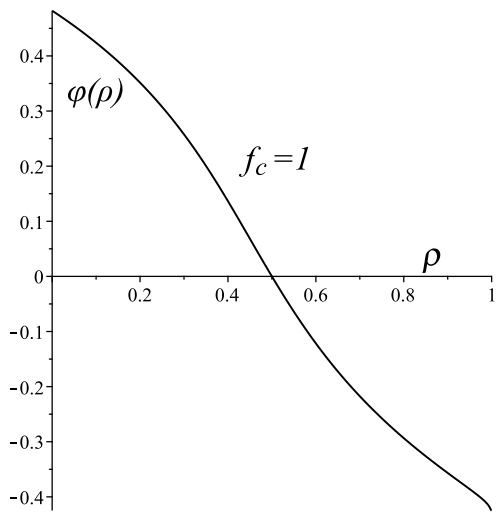}
\caption{\label{fig:ds5} de Sitter solution and a spacelike current vector:  
The solution $\varphi(\rho)$ is plotted in the case $f_c=1$ (black online).
The parameters are fixed as $\Omega=1=H$ and the initial conditions are chosen 
so that $\varphi(0.5)=0$, as an example.}
\end{figure}

\end{enumerate}

\section{Gravitational plane wave pulse}

The last explicit example that we consider is a single pulse of gravitational 
radiation, associated with coordinates $x^{\mu}=(u,v,x,y)$ and described by the metric
\begin{equation}
ds^2=-du dv +\cos^2(u)dx^{2}+\cosh^2(u)dy^{2},
\end{equation}
which solves the vacuum Einstein equations. This specific choice of the metric 
implies $u\in [0,\pi/2]$, with $u=\pi/2$ a coordinate horizon.

\begin{enumerate}
\item $J$ timelike

We choose
\begin{equation}
J=\frac{-2}{\cos u \cosh u}{\partial \over \partial u} 
-2A(u) {\partial \over \partial v}
\Longrightarrow 
J^{\flat} =A(u)du+\frac{1}{\cos u \cosh u}dv,
\end{equation}
so that 
\begin{equation}
\langle J, J \rangle = -\frac{4A(u)}{\cos u \cosh u}.
\end{equation}
Assuming then for the variable $\alpha$ the relation
\begin{equation}
\alpha(u,v)= A(u)(4v+1)^{1/4} ,
\end{equation}
the equation satisfied by $A(u)$ is  
\begin{equation}
\frac{dA}{du}= {1 \over 2}[{\rm tan}(u)-{\rm tanh}(u)]  A(u)
+\frac{f_c^2}{A(u)^2\cos(u)\cosh(u)},
\end{equation}
and can be integrated numerically.

\begin{figure}
\includegraphics[scale=0.35]{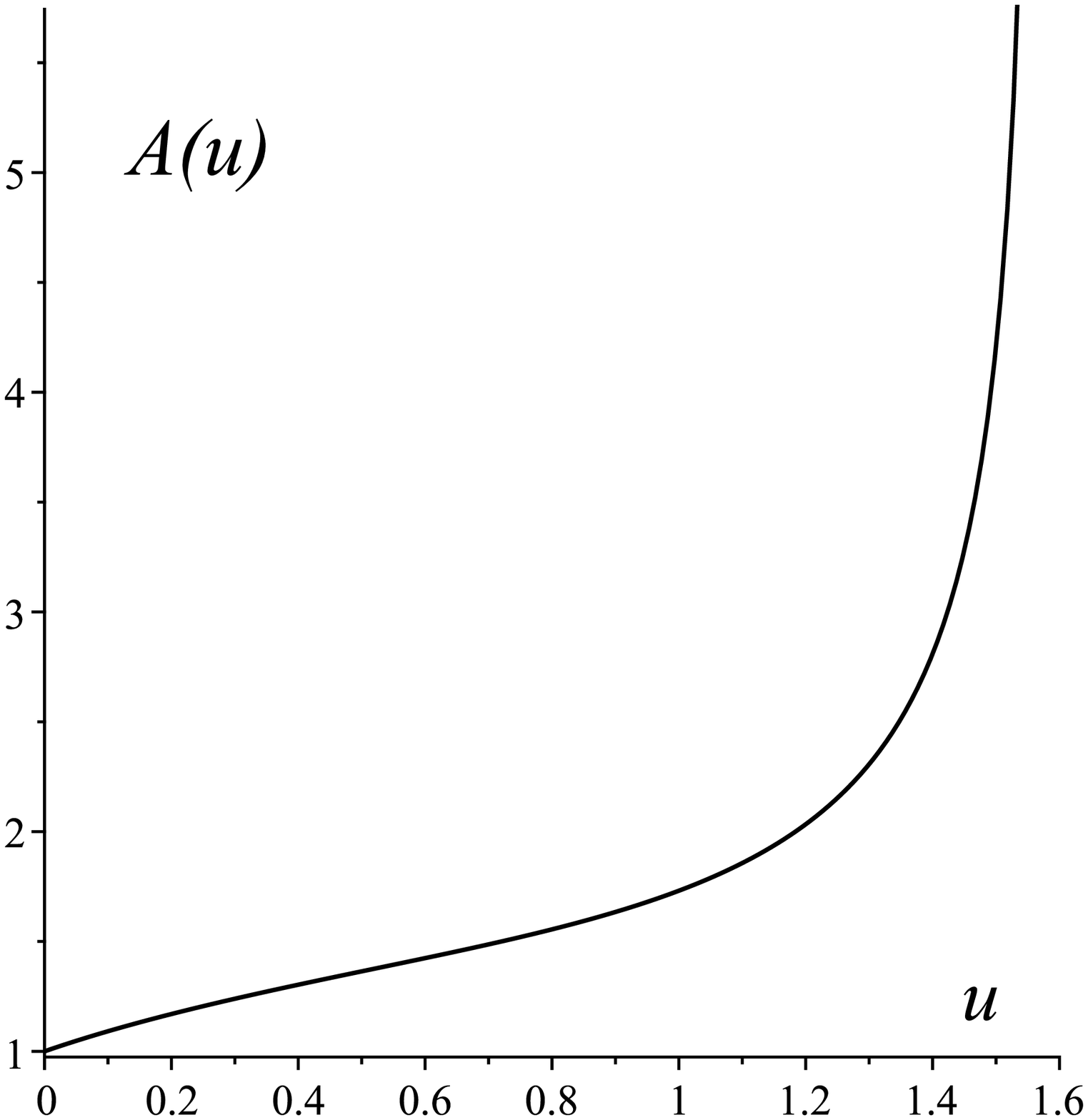}
\caption{\label{fig:6} Gravitational wave solution and a timelike current vector: The solution  
$A(u)$ is plotted in the case $f_c=1$ (black online).
The initial condition is chosen so that $A(0)=1$, as an example.}
\end{figure}

\item $J$ spacelike

We choose
\begin{equation}
J=\frac{A(u)\cos^2 (K_y y)e^{2K_vv}}{\cos^2 u}{\partial \over \partial x} 
\Longrightarrow 
J^{\flat} =A(u)\cos^2 (K_y y)e^{2K_vv} dx 
\end{equation}
so that 
\begin{equation}
\langle J, J \rangle =\frac{A^2(u)\cos^4 (K_y y)e^{4K_vv}}{\cos^2 u}  .
\end{equation}
Assuming then for the variable $\alpha$ the following relation:
\begin{equation}
\alpha(u,v,y)= A(u)\cos(K_{y} y) e^{K_v v}  ,
\end{equation}
the equation satisfied by $A(u)$ is  
\begin{eqnarray}
\frac{dA}{du}
&=&  {1 \over 2} \left[{\rm tan}(u)-{\rm tanh}(u)
-{K_{y}^{2}\over 2 K_{v}\cosh^{2}(u)}\right] A(u)
\nonumber \\&-& 
\frac{f_c^2}{4K_v \cos^{2}(u) A(u)},
\end{eqnarray}
and can be integrated numerically.

\begin{figure}
\includegraphics[scale=0.35]{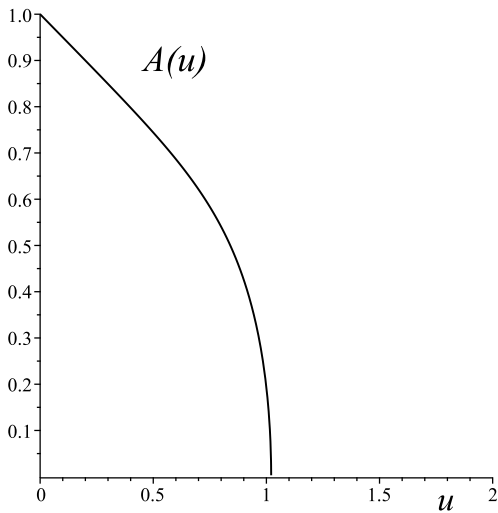}
\caption{\label{fig:6} Gravitational wave solution and a spacelike 
current vector: The solution  
$A(u)$ is plotted in the case $f_c=1$ (black online).
The initial condition is chosen so that $A(0)=1$, as an example.}
\end{figure}

\end{enumerate}

The results of our study indicate qualitatively a blow-up of the field 
at the coordinate horizon $u=\pi/2$.

\section{Concluding remarks}

We have studied an Ermakov-Pinney-like equation  
in curved space-time, elucidating both content and 
role of non-linearities as associated with a divergence-free current source. We have 
discussed the explicit examples of Schwarzschild space-time, de Sitter space-time, and 
the space-time corresponding to single pulse of gravitational radiation in both cases of a 
timelike and spacelike current source, investigating results from the analytic point of 
view, or from a numerical perspective when the analytic solution is not directly available.
The main difficulty of the problem is that of having
to solve the Ermakov-Pinney equation having already 
imposed $J$ to be divergence-free, a fact that poses strong limitations to any general discussion. 
We are forced then to explore special cases and we do this for a few physically relevant 
space-times: black holes (Schwarzschild), cosmological space-times (de Sitter), gravitational wave 
space-time (the metric of a single gravitational wave pulse). 
Of course, one may add similar considerations for other familiar space-times, 
e.g., of cosmological interest like Friedmann-Roberson-Walker 
or exact solutions in presence of matter sources.
However, exploring a larger collection of background space-times is not very 
illuminating, and is not the goal of the present work.
Our aim is to understand (quantitatively, or qualitatively when analytic 
solutions cannot be obtained) the general properties of a massless 
scalar field which interacts with a given gravitational background and is 
also sourced by itself in a (rather simple) non-linear way, a fact which 
complicates matters and leaves most of the analysis to be performed only numerically. 
The Ermakov-Pinney-like equation is then the good candidate we have chosen here.
In all these cases in which the Ermakov-Pinney-like equation occurs we have obtained 
(non-linear) second-order ordinary differential equations
(having already separated the variables or having 
assumed a very simple dependence on the variables) admitting two independent 
solutions with regularity conditions to be imposed properly. 
For example, in the case of a Schwarzschild space-time one can more easily 
find solutions regular at spatial infinity than at the horizon.
Post-Newtonian like solutions also exist and can eventually be combined to achieve
regularity at both points, but this is a not-at-all easy task and 
not the goal of this study.
Indeed, we have \lq\lq preferentially" shown explicit solution which 
are regular at spatial infinity, i.e., where the space-time becomes 
flat and then the interaction of the field with the 
background becomes trivial.
In that case one is left only with the interaction of the field with itself.  
To be more precise:

\begin{enumerate}
\item 
In the Schwarzschild case  
our analysis shows that a positive coupling constant leads to 
spatially damped oscillations of the field, 
whereas a negative or vanishing one is generally associated with blowing-up of the solutions.  
In light of this, in view of the interest in 
studying more thoroughly the quantization properties of fields, it seems  
more relevant to limit considerations to the case of a positive coupling constant, a hot 
topic to be developed in future work. 

It would also be of interest to understand the
relation (if any) with the asymptotic behaviour of solutions of the scalar wave
equation found in Ref. \cite{PRSLA}, as well as with the interesting investigations
performed in Refs. \cite{Casadio,Palia}.

\item
In the de Sitter case, the discussion of analogous situations does not show in general 
oscillations but parabolic-like behaviour, meaning that, from the point of view 
of studying quantization properties of fields, all cases can be  considered.

\item
In the gravitational wave case, the numerical analysis 
shows the occurrence of blow-up behaviours even before reaching the coordinate horizon.

\end{enumerate}

\begin{figure}
\begin{center}
\includegraphics[scale=0.35]{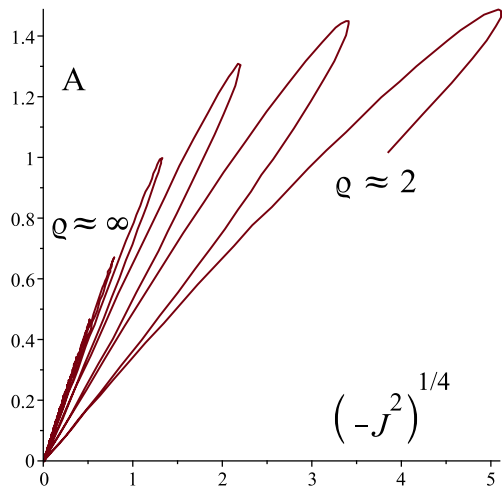}
\caption{\label{fig:11} The case of a Schwarzschild black hole and a timelike 
current $J$ (see Sec. 2.1 above for the description in the text. The numerical 
integration uses the same parameters of Fig. \ref{fig:2}). A  parametric plot 
of $A$ vs $(-\langle J,J\rangle)^{1/4}$ is shown in an attempt to get 
associated coordinate-independent information.}
\end{center}
\end{figure}

As stated above, one can choose initial conditions so as to obtain regular 
solutions not at spatial infinity but at the horizon. This can be done and will 
be addressed in future works. What instead remains to be understood is how to 
extract gauge-invariant informations from these studies. An idea which is 
currently under consideration is to form the magnitude of the current, 
$J^2$, and express the field at any given point not as a function the coordinates  
but of $J^2$ (or of any convenient function of $J^2$). 
For example, in the case of a Schwarzschild black hole and a timelike current 
$J$ (see Sec. 2.1 and Fig. \ref{fig:2} above) a parametric plot of $A(\varrho)$ 
vs $(-\langle J,J\rangle)^{1/4}$ can be seen as a preliminary attempt used to convert 
the above mentioned spatial oscillations in a \lq\lq gauge-invariant" way. 
The result is shown in Fig. \ref{fig:11}, where the right part of the plot 
corresponds to the  horizon while the left part, i.e., the accumulating line,  
corresponds to approaching spatial infinity. The intermediate oscillations 
reflect the oscillating behaviour also seen when working 
in a coordinate-dependent point of view.
This is certainly a possible strategy for expressing gauge-invariant 
information (actually there are no \lq\lq natural" variables in this problem) 
and we are currently analyzing the corresponding limitations and/or advantages, 
besides the effective usefulness. The study in this direction is still preliminary 
and it will remain a challenge for future work. 

\section*{Acknowledgments}

G. Esposito is grateful to the Dipartimento di Fisica ``Ettore Pancini'' of Federico II 
University for hospitality and support.

\end{document}